\documentclass[a4paper]{article}
\usepackage{interspeech2010,amssymb,amsmath,epsfig}
\usepackage{graphicx}
\ninept
\def\reg{{\rm\ooalign{\hfil
     \raise.07ex\hbox{\scriptsize R}\hfil\crcr\mathhexbox20D}}}

\title{Chirp Complex Cepstrum-based Decomposition for Asynchronous Glottal Analysis}

\makeatletter
\def\name#1{\gdef\@name{#1\\}}
\makeatother
\name{{\em Thomas Drugman, Thierry Dutoit}}

\address{TCTS Lab, University of Mons, Belgium \\
}

\begin{document}
\maketitle

\begin{abstract}

It was recently shown that complex cepstrum can be effectively used for glottal flow estimation by separating the causal and anticausal components of speech. In order to guarantee a correct estimation, some constraints on the window have been derived. Among these, the window has to be synchronized on a Glottal Closure Instant. This paper proposes an extension of the complex cepstrum-based decomposition by incorporating a chirp analysis. The resulting method is shown to give a reliable estimation of the glottal flow wherever the window is located. This technique is then suited for its integration in usual speech processing systems, which generally operate in an asynchronous way. Besides its potential for automatic voice quality analysis is highlighted.

%Homomorphic analysis is known to have the ability to linearly separate initially non-linearly combined signals. As a particular case, it has been %discussed that complex cepstrum could be used for source-tract deconvolution, but no-one could achieve this robustly. Relying on the mixed-phase %nature of speech, this paper proposes a study of how to make complex ceptrum decomposition work for glottal flow/vocal tract separation. Emphasize %is put on the crucial role of windowing in order to achieve an accurate deconvolution. A parallel with the Zeros of the Z-Transform (ZZT)-based %decomposition is also drawn. The new technique is shown to yield similar results than ZZT, while it is much faster.
\end{abstract}
\noindent{\bf Index Terms}: Glottal Flow, Voice Quality, Mixed-phase Decomposition.

%% AND NOW START WITH YOUR PAPER CONTENT

%%%%%%%%%%%%%%%%%%%%%%%%%%%%%%%%%%%%%%%%%%%%%%%%%%%%%%%%%%%%%%%%%%
\section{Introduction}\label{sec:intro}
%%%%%%%%%%%%%%%%%%%%%%%%%%%%%%%%%%%%%%%%%%%%%%%%%%%%%%%%%%%%%%%%%%

Complex Cepstrum has recently shown its ability to efficiently estimate the glottal flow \cite{Drugman-CCD}. An essential constraint for leading to a correct source-tract separation with this technique is the condition of being synchronized on Glottal Closure Instants (GCIs). On the other hand, the large majority of current speech processing systems operate in an asynchronous way, i.e use a constant frame shift. This paper proposes a modification of the complex cepstrum-based decomposition so as to integrate this technique in such systems.

The paper is structured as follows. Section \ref{sec:CCD} reviews the principles of the Complex Cepstrum-based Decomposition (CCD). Section \ref{sec:Chirp} extends this formalism to integrate a chirp analysis. Besides an automatic way to find the optimal chirp contour is proposed. In Section \ref{sec:results} the performance of the resulting method is evaluated on a large corpus of emotional speech. It is shown that the chirp CCD technique gives a reliable asynchronous estimation of the glottal flow and can be used for voice quality analysis. Finally Section \ref{sec:conclu} concludes.

%%%%%%%%%%%%%%%%%%%%%%%%%%%%%%%%%%%%%%%%%%%%%%%%%%%%%%%%%%%%%%%%%%
\section{Complex Cepstrum for Glottal Source Estimation}\label{sec:CCD}
%%%%%%%%%%%%%%%%%%%%%%%%%%%%%%%%%%%%%%%%%%%%%%%%%%%%%%%%%%%%%%%%%%

%In \cite{Drugman-CCD}, we showed that Complex Cepstrum can be efficiently used for glottal source estimation. 
The principle of the Complex Cepstrum-based Decomposition (CCD, \cite{Drugman-CCD}) relies on the mixed-phase model of speech \cite{MixedPhase}. According to this model, speech is composed of both minimum-phase (i.e causal) and maximum-phase (i.e anticausal) components. While the vocal tract and the glottal \emph{return phase} can be considered as minimum-phase systems, it has been shown \cite{Doval-CALM} that the glottal \emph{open phase} is a maximum-phase signal. The key idea of the mixed-phase decomposition is then to separate both minimum and maximum-phase components of speech, where the latter is only due to the glottal contribution. In previous works, we proposed two algorithms achieving the mixed-phase decomposition: the Zeros of the Z-Transform (ZZT) algorithm \cite{ZZT}, and the Complex Cepstrum-based Decomposition (CCD, \cite{Drugman-CCD}). Although both techniques are functionnaly equivalent, CCD was shown \cite{Drugman-CCD} to be much faster than ZZT. For this reason, this paper only focuses on the use of the Complex Cepstrum.

The Complex Cepstrum (CC) $\hat{x}(n)$ of a discrete signal $x(n)$ is defined by the following equations \cite{Oppenheim}:

\begin{equation}\label{eq:DFT}
X(\omega)=\sum_{n=-\infty}^{\infty} x(n)e^{-j\omega n}
\end{equation}

\begin{equation}\label{eq:ComplexLog}
\log[X(\omega)]=\log(|X(\omega)|)+j\angle{X(\omega)}
\end{equation}

\begin{equation}\label{eq:ComplexCepstrum}
\hat{x}(n)=\frac{1}{2\pi}\int_{-\pi}^{\pi}{\log[X(\omega)]e^{j\omega n}}\emph{d}\omega
\end{equation}

where Equations \ref{eq:DFT}, \ref{eq:ComplexLog}, \ref{eq:ComplexCepstrum}  are respectively the Discrete-Time Fourier Transform (DTFT), the complex logarithm and the inverse DTFT (IDTFT). Our decomposition arises from the fact that the complex cepstrum $\hat{x}(n)$ of an anticausal (causal) signal is zero for all $n$ positive (negative). Retaining only the negative indexes of the CC should then estimate the glottal open phase contribution. CCD consequently achieves the mixed-phase decomposition using the quefrency origine as a discriminant.

Nonetheless it was shown in \cite{Drugman-CCD} that windowing is crucial and dramatically conditions the efficiency of the method. It is indeed essential that the window applied to the voiced segment of speech $x(n)$ respects some constraints in order to exhibit correct mixed-phase properties. Among these constraints, the window should \cite{Drugman-CCD}:

\begin{itemize}
\item be synchronized on a Glottal Closure Instant (GCI),

\item have an appropriate shape and length (proportional to the pitch period).
\end{itemize}

Although some works aim at estimating the GCI positions directly from the speech signal \cite{Drugman-GCI}, or use ElectroGlottoGraphs (EGGs, \cite{EGG}), the great majority of current speech processing systems do not have this information available and consequently operate in an asynchronous way. In the following Section, we extend the Complex Cepstrum formalism so as to remove the GCI-synchronization constraint and propose an automatic way to achieve this.

%%%%%%%%%%%%%%%%%%%%%%%%%%%%%%%%%%%%%%%%%%%%%%%%%%%%%%%%%%%%%%%%%%
\section{Extension to the Chirp Analysis}\label{sec:Chirp}
%%%%%%%%%%%%%%%%%%%%%%%%%%%%%%%%%%%%%%%%%%%%%%%%%%%%%%%%%%%%%%%%%%
The Chirp Z-Transform (CZT), as introduced by Rabiner et al \cite{chirp} in 1969, allows the evaluation of the z-transform on a spiral contour in the z-plane. Its first application aimed at separating too close formants by reducing their bandwidth. In a previous work \cite{Drugman-Chirp}, we showed for the Zeros of the Z-Transform (ZZT) algorithm that considering a contour possibly different from the unit circle makes the method more robust to GCI location errors. In its original version \cite{ZZT}, the ZZT technique achieves the mixed-phase decomposition by separating the roots of the polynomial $X(z)$ located inside and outside the unit circle in the z-plane (respectively corresponding to the minimum and maximum-phase components of speech). The key idea of the work described in \cite{Drugman-Chirp} was then to evaluate the CZT on a circle of radius $R$ (with possibly $R\neq 1$) such that the root distribution is split into two well-separated groups. More precisely, it was observed that the significant impulse present in the excitation at the GCI results in a gap in the root distribution. When analysis is exactly GCI-synchronous, the unit circle perfectly separates causal and anticausal roots. On the opposite, when the window moves off from the GCI, the root distribution is transformed. Such a decomposition is then not guaranteed for the unit circle and another boundary is generally required.

\begin{figure}[!ht]
  \centering
  \includegraphics[width=0.45\textwidth]{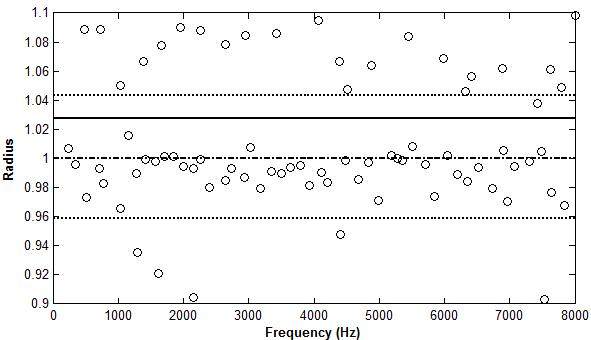}
  \caption{Representation of the Zeros of the Z-Transform in polar coordinates. The optimal chirp circle (solid line) gives the best separation of the root distribution within the bounds $exp(\pm\frac{50\pi}{17L})$ (dotted lines) and leads to a correct source-filter separation. It is clearly seen that, in this case, the unit circle (dashdotted line) will lead to an erroneous decomposition.}
  \label{fig:ChirpZZT}
\end{figure}

Figure \ref{fig:ChirpZZT} gives an example of root distribution for a natural voiced speech frame for which a timing error is made on the actual GCI position. It is clearly seen that using the traditional z-transform ($R=1$) for this frame will lead to an erroneous decomposition. In contrast, it is possible to find an optimal radius leading to a correct separation, as indicated with a solid line in Figure \ref{fig:ChirpZZT}. In \cite{Drugman-Chirp} it was also demonstrated that, for a two pitch period-long Blackman window (satisfying the second constraint of Section \ref{sec:CCD}), the optimal radius is comprised within the bounds $exp(\pm\frac{50\pi}{17L})$ where $L$ denotes the frame length in samples (these bounds are indicated in Figure \ref{fig:ChirpZZT} in dotted lines).

Relying on these conclusions, it is here proposed to integrate the chirp analysis whitin the Complex Cepstrum-based Decomposition (CCD) and to automatically find the optimal circle \emph{without requiring the computation of the root distribution} (so as to still benefit from the speed of the CCD algorithm).

Achieving a chirp ZZT-based decomposition is straightforward since it is only necessary to modify the radius used to isolate the maximum-phase component. In order to integrate the chirp analysis for the CCD technique, let us consider the signal $x(n)$. Its CZT evaluated on a circle of radius $R$ can be written as \cite{chirp}:

\begin{align}
X(Rz)&=\sum_{n=0}^{L-1}x(n)(Rz)^{-n}=\sum_{n=0}^{L-1}(x(n)R^{-n})z^{-n}
\end{align}

and is consequently equivalent to evaluating the z-transform of a signal $x_R'(n)=x(n)R^{-n}$ on the unit circle. The chirp CCD computed on a circle of radius $R$ can therefore be achieved by applying the traditional CCD framework described in Section \ref{sec:CCD} to $x_R'(n)$ instead of $x(n)$.

In order to automatically estimate the radius giving an optimal separation between minimum and maximum-phase contributions, the unwrapped phase $\phi_R'(\omega)$ of $x_R'(n)$ is inspected. More precisely, the radius axis is uniformly discretized in $N$ values ($N=60$ in our experiments) between the bounds $exp(\pm\frac{50\pi}{17L})$. For each radius value $R$, $\phi_R'(\omega)$ is computed and the linear phase component is characterized by $\phi_R'(\pi)$ (with $\phi_R'(0)=0$ by definition). From this, we define the variable $n_d(R)$:

\begin{align}
n_d(R)=\frac{\phi_R'(\pi)}{\pi}
\end{align}

as the number of samples of circular delay, i.e the number of samples that $x_R'(n)$ shoud be circularly shifted so as to remove its linear phase. 

%From this, we define the variable $n_d(R)=\frac{\phi_R'(\pi)}{\pi}$ as the number of samples of circular delay, i.e the number of samples that $x_R'(n)$ shoud be circularly shifted so as to remove its linear phase. 

\begin{figure}[!ht]
  \centering
  \includegraphics[width=0.45\textwidth]{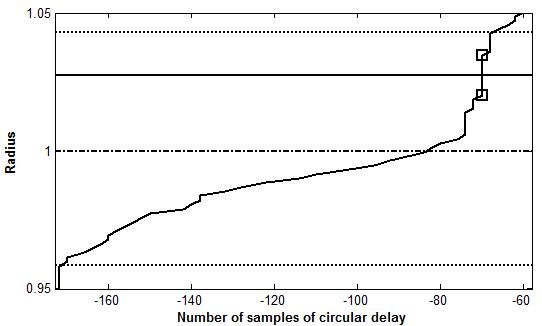}
  \caption{Evolution of $n_d(R)$ for the same signal as in Figure \ref{fig:ChirpZZT}. The optimal radius (solid line) is defined as the middle of the largest interval (indicated by squares) for which $n_d(R)$ stays constant, within the bounds $exp(\pm\frac{50\pi}{17L})$ (dotted lines). The unit circle used in the traditional CCD is represented in dashdotted line. }
  \label{fig:ChirpCCD}
\end{figure}

\begin{figure}[!ht]
  \centering
  \includegraphics[width=0.25\textwidth]{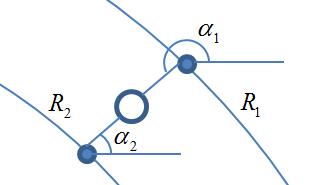}
  \caption{Illustration of a phase jump of $-\pi$ due to the passage of a zero from the inside of the circle of radius $R_1$ to the outside of the circle of radius $R_2$.}
  \label{fig:PhaseJump}
\end{figure}

Figure \ref{fig:ChirpCCD} shows the evolution of $n_d(R)$ for the same signal as used in Figure \ref{fig:ChirpZZT}. $n_d(R)$ is actually a step function where gaps are due to the passage of some roots from the inside to the outside of the considered chirp circle. These phase discontinuities are illustrated in Figure \ref{fig:PhaseJump}. Indeed consider a zero which, initially located inside the circle of radius $R_1$ used for the evaluation of the CZT, is now passed outside of the circle of radius $R_2$ (with $R_1>R_2$). When the CZT is evaluated on a point close to this zero in the z-plane, this results in a phase jump of $-\pi$ (see angles $\alpha_1$ and $\alpha_2$ in Fig. \ref{fig:PhaseJump}) which is then reflected in $\phi_R'(\pi)$. The difference $n_d(R_1)-n_d(R_2)$ is consequently interpreted as the number of zeros which, initially inside the circle of radius $R_1$, have passed the boundary to be now located outside the circle of radius $R_2$.
In other words, inspecting the variable $n_d(R)$ allows us to detect the discontinuities in the root distribution (without requiring its whole computation). Similarly to \cite{Drugman-Chirp}, the optimal radius used for the chirp CCD is then defined as the middle of the largest interval for which $n_d(R)$ stays constant, within the bounds $exp(\pm\frac{50\pi}{17L})$. 

%%%%%%%%%%%%%%%%%%%%%%%%%%%%%%%%%%%%%%%%%%%%%%%%%%%%%%%%%%%%%%%%%%
\section{Experiments}\label{sec:results}
%%%%%%%%%%%%%%%%%%%%%%%%%%%%%%%%%%%%%%%%%%%%%%%%%%%%%%%%%%%%%%%%%%
Experiments are carried out on the De7 corpus. This database was designed by Marc Schroeder as one of the first attempts of creating diphone databases for expressive speech synthesis \cite{Schroder}. The database contains three voice qualities (modal, soft and loud) uttered by a German female speaker, with about 50 minutes of speech available for each voice quality. Besides, GCI positions are estimated by the algorithm described in \cite{Drugman-GCI}.

The goal of this Section is to compare the traditional and the proposed chirp CCD techniques by studying their efficiency for glottal source estimation. Experiments are divided into two parts. In the first one, the sensitivity of both methods to GCI location errors is investigated. In the second part, the whole expressive speech database is analyzed by the two techniques and it is shown that chirp CCD leads to results similar as with the traditional CCD, but without the requirement of operating in a GCI-synchronous way.

\subsection{Robustness to GCI location errors}\label{ssec:Robustness}
 
\begin{figure}[!ht]
  \centering
  \includegraphics[width=0.45\textwidth]{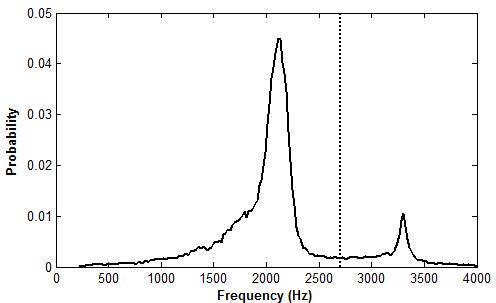}
  \caption{Distribution of the spectral center of gravity of the maximum-phase component, computed for the whole dataset of loud utterances. Fixing a threshold around 2.7kHz makes a good separation between correctly and incorrectly decomposed frames.}
  \label{fig:HistoCOG}
\end{figure}

When performing the mixed-phase separation, it may appear for some frames that the decomposition is erroneous, leading to an irrelevant high-frequency noise in the estimated glottal source \cite{ZZT}. As a criterion deciding whether a frame is considered as correctly decomposed or not, the spectral center of gravity is inspected. The distribution of this feature is displayed in Figure \ref{fig:HistoCOG} for the loud voice using the traditional CCD. A principal mode at around 2kHz clearly emerges and corresponds to the majority of frames for which a correct decomposition is carried out. A second minor mode at higher frequencies is also observed. It is related to the frames where the mixed-phase decomposition fails, leading to a maximum-phase signal containing an irrelevant high-frequency noise. It can be noticed from this histogram (and it was confirmed by a manual verification of numerous frames) that fixing a threshold at around 2.7kHz makes a good distinction between frames that are correctly and incorrectly decomposed.

\begin{figure}[!ht]
  \centering
  \includegraphics[width=0.45\textwidth]{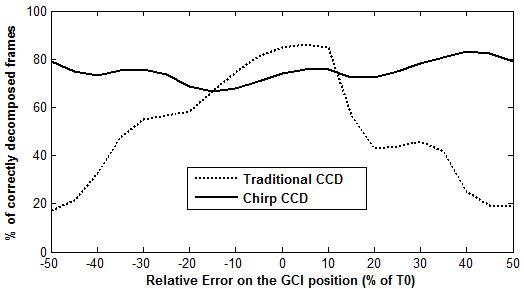}
  \caption{Robustness of both traditional and chirp CCD methods to a timing error on the GCI location.}
  \label{fig:ShiftInfluence}
\end{figure}

Given this criterion, the sensitivity of both traditional and chirp CCD techniques to a GCI location error is displayed in Figure \ref{fig:ShiftInfluence} for the loud dataset of the De7 corpus. The constraint of being GCI-synchronous for the traditional CCD is clearly confirmed on this graph. It is indeed seen that the performance dramatically degrades for this technique as the window center moves off from the GCI. On the contrary, the chirp CCD method gives a high rate of correctly decomposed frames (however slightly below the performance of the GCI-centered traditional CCD) wherever the window is located.

\subsection{Asynchronous glottal analysis of emotional speech}

In this Section, we confirm the potential of the chirp CCD technique for asynchronously estimating the glottal flow on a large speech corpus. For this, the whole De7 database with its 3 voice qualities is analyzed. The glottal flow is estimated by 2 techniques:

\begin{itemize}
\item {the traditional CCD:} analysis is GCI-synchronous,
\item {the chirp CCD:} analysis is asynchronous. A constant frame shift of $10ms$ is considered, as widely used in many speech processing systems. Note however that a two pitch period-long Blackman window is applied, as this is essential for achieving a correct mixed-phase decomposition (see Section \ref{sec:CCD}).
\end{itemize}

In a first time, we evaluate the proportion of frames that are correctly decomposed by these two techniques using the spectral criterion of Section \ref{ssec:Robustness}. Overall results for the three voice qualities are summarized in Table \ref{tab:TabCorrect}. The traditional CCD gives relatively high rates of correct decomposition with around $85\%$ for the three datasets. It can also be observed that the chirp CCD method makes double the erroneous decompositions than the traditional approach. Nevertheless a correct estimation of the glottal source is carried out by the chirp CCD for around $70\%$ of speech frames, which is rather high for real connected speech. 

\begin{table}[!ht]
\centering
\begin{tabular}{| c || c | c | c |}
\hline
 Method & Loud & Modal & Soft\\
\hline
traditional CCD & 87.09 & 84.41 & 83.68\\
\hline
chirp CCD & 76.43 & 68.07 & 67.48\\
\hline
\end{tabular}
\caption{Proportion ($\%$) of correctly decomposed frames using the traditional and the chirp CCD techniques for the three voice qualities of the De7 database.}
\label{tab:TabCorrect}
\end{table}

\begin{figure*}[!ht]
  \centering
  \includegraphics[width=0.95\textwidth]{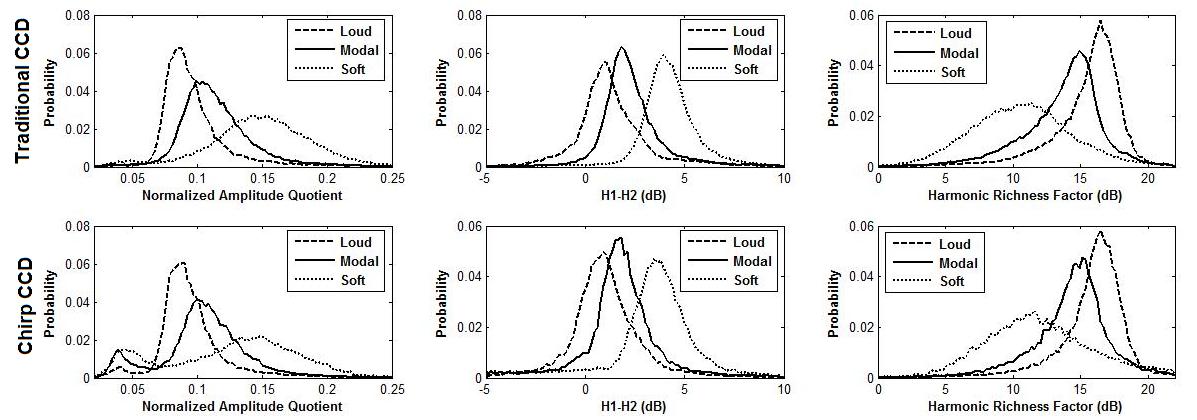}
  \caption{Distributions of glottal parameters estimated by (from top to bottom) the traditional and chirp CCD techniques, for three voice qualities. The considered glottal features are (from left to right): the Normalized Amplitude Quotient (NAQ), the H1-H2 ratio and the Harmonic Richness Factor (HRF).}
  \label{fig:HistosParams}
\end{figure*}

In a second time, frames of the glottal flow that were correctly estimated are characterized by the three following features: the Normalized Amplitude Quotient (NAQ), the H1-H2 ratio between the two first harmonics and the Harmonic Richness Factor (HRF). These glottal parameters were shown in \cite{Alku-NAQ} and \cite{Alku-CP} to lead to a good separation between different types of phonation. The histograms of these parameters estimated by both traditional and chirp CCD methods are displayed in Figure \ref{fig:HistosParams} for the three voice qualities. Two main conclusions can be drawn from this Figure. First, it turns out that the distributions obtained by both techniques are strongly similar. A minor difference can however be noticed for NAQ histograms, where the distributions obtained by the chirp method contain a weak irrelevant peak at low NAQ values. The second important conclusion is that both techniques can be efficiently used for glottal-based voice quality analysis, leading to a clear discrimination between various phonation types.

%%%%%%%%%%%%%%%%%%%%%%%%%%%%%%%%%%%%%%%%%%%%%%%%%%%%%%%%%%%%%%%%%%
\section{Conclusion}\label{sec:conclu}
%%%%%%%%%%%%%%%%%%%%%%%%%%%%%%%%%%%%%%%%%%%%%%%%%%%%%%%%%%%%%%%%%%
This paper proposed an extension of the traditional Complex Cepstrum-based Decomposition (CCD). For this, the z-transform was evaluated on a contour in the z-plane possibly different from the unit circle. Circular contours were considered and an automatic way to find the optimal radius leading to well-separated groups of zeros was proposed. The resulting method was shown to be much more robust to GCI location errors than the traditional CCD approach. More particularly a reliable estimation of the glottal flow was obtained in an asynchronous way on real connected speech. Besides this technique showed its potential to be used for automatic voice quality analysis.

%%%%%%%%%%%%%%%%%%%%%%%%%%%%%%%%%%%%%%%%%%%%%%%%%%%%%%%%%%%%%%%%%%
\section{Acknowledgments}\label{sec:Acknowledgments}
%%%%%%%%%%%%%%%%%%%%%%%%%%%%%%%%%%%%%%%%%%%%%%%%%%%%%%%%%%%%%%%%%%

Thomas Drugman is supported by the ``Fonds National de la Recherche
Scientifique'' (FNRS). Authors are also thankful to M. Schroeder for providing the De7 database.

\eightpt
\bibliographystyle{IEEEtran}

\end{document}